# Rate of Excitation Energy Transfer between Fluorescent Dyes and Nanoparticles

Sangeeta Saini,[1] Somnath Bhowmick,[2] Vijay B. Shenoy,[2] Biman Bagchi[1,*]

[1]Solid State and Structural Chemistry Unit, Indian Institute of Science, Bangalore 560 012, India.

[2]Materials Research Center, Indian Institute of Science, Bangalore 560 012, India.

## ABSTRACT

Long range resonance energy transfer (RET) between a donor and an acceptor molecule is increasingly being used in many areas of biological and material science. The phenomenon is used to monitor the in vivo separation between different (bio) polymers/units of (bio) polymers and hence the dynamics of various biomolecular processes. Because of the sensitivity of the rate on to the distance between the donor (D) and the acceptor (A), the technique is popularly termed as "spectroscopic ruler". In this work we examine the distance and orientation dependence of RET in three different systems: (i) between a conjugated polymer and a fluorescent dye, (ii) between a nanometal particle (NMP) and a fluorescent dye and (iii) between two NMP. We show that in all the three cases, the rate of RET follows a distance dependence of $d^{-\sigma}$ where exponent $\sigma$ approaches 6 at large $d$ (Förster type dependence) but has a value varying from 3 - 4 at short to intermediate distance. At short separation, the amplitude of rate is considerably smaller than predicted by the Förster theory.

**Corresponding Author:** Tel.: +91 80 22932926; fax: +91 80 23601310
*E-mail:* bbagchi@sscu.iisc.ernet.in

# 1. Introduction

Resonance energy transfer (RET) is a widely prevalent photophysical process through which an electronically excited 'donor' molecule transfers its excitation energy to an 'acceptor' molecule such that the excited state lifetime of the donor decreases.[1, 2] If the donor happens to be a fluorescent molecule RET is referred to as fluorescence resonance energy transfer, FRET, although the process is non-radiative. The acceptor may or may not be fluorescent. Energy conservation requires that the energy gaps between the ground and the excited states of participating donor and acceptor molecules are nearly the same. This in turn implies that the fluorescence emission spectrum of the donor (D) must overlap with the absorption spectrum of the acceptor (A), and the two should be within the minimal spatial range for donor to transfer its excitation energy to the acceptor.

In early 1920's the fluorescence quenching experiments revealed the phenomenon of FRET and led J. Perrin[3] to propose dipole-dipole interaction as a mechanism via which molecules can exchange energy over distances much greater than their molecular diameter. Later on, Förster[1] built upon Perrin's idea to put forward an elegant theory which provided a quantitative explanation for the non-radiative energy transfer, given by

$$k_F = k_{rad} \left( R_F / R \right)^6 \qquad (1)$$

where $k_{rad}$ is the radiative rate (typically less than $10^9$ s$^{-1}$) and $R_F$ is the well-known Förster radius given by the spectral overlap between the fluorescence spectrum of the donor and the absorption spectrum of the acceptor. Since then the technique of FRET has come a long way finding applications in most of the disciplines (chemistry, biology and material science). It is often designated as a "spectroscopic ruler"[4] because the strong distance dependence of the energy transfer rate provides us with a microscopic scale to measure separations in vivo, typically in range of 20 – 80 Å. Undoubtedly, FRET has played a key role in understanding the conformational dynamics of single (bio) molecules in microscopic detail.[5-8] However, the conventional FRET (both donor and acceptor are dye molecules) suffers from several limitations prominent among them is the restriction on the upper limit of separation of only 80 Å. Beyond this distance, the energy transfer becomes too weak to be useful.[9] This limitation has motivated the use of RET systems involving dye molecules and the noble metal nanoparticles. These nanoparticles (Au, Ag) have prominent absorption spectrum in the visible region and are employed as either the acceptor, or more recently as both donor and the acceptor.[10] The absorption of light by nanoparticles is mainly dominated by the surface plasmon (SP) resonance.[11, 12] In such RET systems, separations up to 700 Å can be monitored. This feature makes these RET systems potentially extremely useful in many material and biomedical applications.[13]

In the present paper we will discuss the distance and orientation dependence of energy transfer between different donor-acceptor systems. In section II, we will focus upon the excitation energy transfer (EET) between conjugated dye molecules, the conventional RET system. In section III, we will consider the EET from a dye to a metal nanoparticle and in section IV we will briefly discuss resonance energy transfer between two nanoparticles.

## 2. Excitation energy transfer in conjugated systems

EET is a basic function of photosynthetic antennas which collect and channel the harvested solar energy to reaction center with about 95% efficiency.[14] These efficient light harvesting systems are nothing but extensively conjugated organic systems. As a consequence the process of EET in conjugated systems is being foreseen as a mode of signal transmission in molecular electronics. Moreover, these systems are already finding number of applications in display devices.[15, 16]

To design the systems as efficient as photosynthetic antennas, the first step is to direct the flow of energy towards the desired regions. To attain this objective scientists have linked chromophores with continually decreasing band gaps along the polymer chains.[17] However, in such systems energy is lost because of large differences required in the emission spectra of successive chromophores.

A novel system that controls the energy flow more efficiently has been designed by Schwartz *et al.*[18] It involve the chains of semiconducting poly[2-methoxy-5-(2'-ethyl-hexyloxy)-1,4-phenylene vinylene] (MEH-PPV) aligned and encapsulated into the hexagonally arrayed channels of mesoporous silica glass. PPV and hence its water soluble derivative, MEH-PPV is not infinitely long conjugated system but consists of chains of conjugated phenyl-vinyl oligomers of various lengths because of defects, bends and twists in polymer chain. These oligomers can serve both as acceptor and donor molecules in non-radiative excitation energy transfer as excitation energy is a strong function of conjugation length.

In this system, the polymer outside the channel is composed of short randomly oriented oligomer having high excitation energy while that inside oriented along the channel consist of longer segments. This particular design directs the energy deposited with the randomly oriented segments towards the aligned ones inside the channels. With the help of both steady-state and time-resolved luminescence measurements Schwartz *et al.* concluded that the dominant interchain energy transfer (energy transfer between the randomly oriented chains outside the channels) mechanism as the Förster energy transfer.

Since the interchain migration rate depends on the relative internal geometries of the donor and acceptor chromophores, an understanding of spatial and orientation dependence of the rate of excitation energy transfer is therefore important for optimizing the performance of molecular-based devices involved in EET. Wong *et al.*[19] have investigated this dependence for a six-unit oligomer of polyfluorene ($PF_6$) and tetraphenylporphyrin (TPP) which brought forward several limitations of Förster theory. The representative orientations and the structures of two polymers are given in Figure 1. The computational approach employed semiempirical Pariser-Parr-Pople (PPP) hamiltonian coupled with single configuration interaction (SCI). From the PPP/SCI wave functions, electronic transition energies, and the transition dipole moments, the full resonance-Coulomb coupling matrix elements as well as the point-dipole approximation to the coupling were computed.

The comparison of the calculated distance and orientation dependence of Förster rate to the full resonance-Coulomb rate from identical wave functions clearly delineated the limitations of the point-dipole formulation, which is invalid at short D-A separations. The plot of the rate dependence of EET between the donor state and the acceptor state against the DA separation (Fig.2) clearly indicates the violation of Förster distance dependence at small DA separations. This difference from the Förster's macroscopic formulation is a manifestation of the breakdown of the point-dipole approximation. Also the plot shows that the transitions to the acceptor states having mid-range oscillator strength (0.70) dominate the total rate. The separate calculation of transfer rate between optically dark states of D and A shows the rate to be of same order as that between the optically bright states suggesting that while these optically dark states do not contribute to absorption spectrum of the acceptor, they can mediate EET.

Figure 3 shows the orientation dependence for the cofacial case (Fig.1) for two DA separation distances, 10 Å and 100 Å. Here, the angle $\theta$ corresponds to the rotation of TPP acceptor molecule about the transition dipole moment axis (z-axis in Fig.1). Fig. 3a shows the rate to vary by a factor of ~2 in going from 0° to 90° whereas the dipole approximation to the rate shows the negligible dependence. However, at large separations (Fig. 3b) the EET rate shows weak orientation dependence.

In brief, the study by Wong *et al*. proves the inadequacy of the Förster theory at short distances particularly for extended conjugated system where the transition dipole densities are distributed on the length scale similar to DA separation.

## 3. Excitation energy transfer from a dye to a metal nanoparticle

In the vicinity of a metal surface both the radiative lifetime of the dye molecule and the rate of non-radiative energy transfer changes. A large number of theoretical and experimental studies exist on the rate of non-radiative energy transfer from a dye to both, a plane metallic surface[20-22] and a nanoparticle.[23-27] However, only a few of the studies explore the distance dependence of non-radiative energy transfer. In present section, we will discuss the dependence of the rate on the separation between a dye and a nanoparticle, the orientation of the dye molecule with respect to the distance vector ***R*** (Fig. 4) and the size of the nanoparticle.

First we briefly discuss a quantum mechanical approach adapted by us to calculate the non-radiative decay rate of a dye molecule in presence of a metallic nanoparticle.[28] The formalism differs from the earlier theoretical studies in being more microscopic which appeals directly to the collective electronic excitations of the nanoparticle. Since the absorption spectrum of metal nanoparticles over a wide range of size is dominated by surface plasmon resonance, the formalism invokes the transfer of excitation energy to the surface plasmon modes of the nanoparticle.

According to Fermi golden rule, the rate of energy transfer is given by

$$k_{DA} = \frac{2\pi}{\hbar}|V_{DA}|^2 \sum_{M_D,N_D}\sum_{M_A,N_A} f(E^e_{N_D})\ f(E^g_{M_A})\left|\left\langle \chi^g_D;\chi^e_A\middle|\chi^e_D;\chi^g_A\right\rangle\right|^2 \times$$
$$\delta(E^e_{N_D} + E^g_{M_A} - E^g_{M_D} - E^e_{N_A}) \quad (2)$$

where

$$V_{DA} = \left\langle \psi^g_{M_D};\psi^e_{N_A}\middle|H_I\middle|\psi^e_{N_D};\psi^g_{M_A}\right\rangle \quad (3)$$

Here $H_I$ is the interaction Hamiltonian and $\left\langle \chi^g_D;\chi^e_A\middle|\chi^e_D;\chi^g_A\right\rangle$ are the matrix elements of nuclear overlap factors, and $\psi$ denotes the electronic wave function. The delta function satisfies the condition of energy conservation. The sum is over all the possible vibrational states of donor molecule and the various other degrees of freedom of nanoparticle (like the interaction with phonons, electron-hole pair interactions which broadens the absorption spectrum of the nanoparticle) weighed by their initial thermal distribution, $f(E^e_{N_D})$ and $f(E^g_{M_A})$, respectively. The delta function in Eq. (2) can be written as,

$$\delta(E^e_{N_D} + E^g_{M_A} - E^g_{M_D} - E^e_{N_A}) = \int_{-\infty}^{\infty} dE\ \delta(E^e_{N_D} - E^g_{M_D} - E) \times$$
$$\delta(E^g_{M_A} - E^e_{N_A} + E) \quad (4)$$

In the theoretical implementation of the scheme, the sharp resonance lines are replaced with Lorentzians of width 0.025 eV in order to account for the broadening caused by various degrees of freedom.

We model the dye molecule as a particle in a box. The corresponding charge density operator in terms of creation $(c^\dagger_m)$ and annihilation $(c_m)$ operators is given by

$$\widehat{\rho_d}(\vec{r}_D) = -\frac{2e}{L}\sum_{m,n}\psi^*_m(\vec{r}_D)\psi_n(\vec{r}_D)c^\dagger_m c_n + \frac{e}{L}\sum_p c^\dagger_m c_m \quad (5)$$

where $\psi$s are the electronic wavefunctions which depend on $\vec{r}_D$, the position vector of a point in the dye measured from its center, $e$ is the magnitude of electron charge and $2L$ is the length of the 1D box. The second term accounts for the uniformly distributed positive charge background and ensures the overall charge neutrality of the dye molecule. The charge density operator for the nanoparticle within the electrohydrodynamic approximation is given by

$$\rho(\vec{r}_A) = \sum_{l,m} \rho_{l,m}(\alpha \vec{r}_A) = \sum_{l,m} A_{l,m} j_l(\alpha \vec{r}_A) Y_{l,m}(\theta,\varphi) \tag{6}$$

where $Y_{l,m}$ represent the spherical harmonics and $A_{l,m}$ is the amplitude operator given in terms of the plasmon bosonic operators as

$$A_{l,m} = \left( \frac{\hbar \omega_{l,m} \varepsilon_0 \alpha^2}{2a^3} \right)^{1/2} (a^\dagger_{l,m} + a_{l,m})$$
$$\left[ \frac{\omega^2_{l,m}}{\omega^2_p} \frac{(2l+1)}{2} j^2_{l+1}(\alpha a) - j_{l-1}(\alpha a) j_{l+1}(\alpha a) \right]^{-1/2} \tag{7}$$

We study the rate of energy transfer between dye and the nanoparticle using two different interaction Hamiltonian. The full Coulombic interaction Hamiltonian is given by

$$H_I = \frac{1}{4\pi\varepsilon_0} \int d\vec{r}_D \int d^3\vec{r}_A \frac{\rho(\vec{r}_A)\rho(\vec{r}_D)}{\left| \vec{R} - \vec{r}_A + \vec{r}_D \right|} \tag{8}$$

while the interaction Hamiltonian within the dipole approximation is given by

$$H_I = \frac{1}{4\pi\varepsilon_0} \left[ \frac{\vec{\mu}_D \cdot \vec{\mu}_A - 3(\vec{\mu}_D \cdot \hat{d})(\vec{\mu}_A \cdot \hat{d})}{d^3} \right] \tag{9}$$

where $\vec{\mu}_D$ and $\vec{\mu}_A$ are the dipole operators of the dye and the nanoparticle respectively. $\vec{d}$ is the distance between dye and the surface of the nanoparticle and $\hat{d}$ is the corresponding unit vector (see Fig. 4)

The rate of energy transfer is calculated for a donor dye molecule emitting at 520 nm. The acceptor is a nanoparticle with plasma frequency $\omega_p = 5.7 \times 10^{15} s^{-1}$ (after Ref. 29; which fixes the value of the electron density $n_0$). The plasmon frequencies have strong size dependence for particles with $a \leq 7nm$, and asymptotically reaches a plateau value (independent of the size of nanoparticle) for larger particles. The frequencies of the surface modes are lower than $\omega_p$; for a nanoparticle of radius $\sim 10$ $nm$, the $l=1$ surface plasmon has an absorption spectrum centered around 590 nm, while the $l=2$ mode is centred around 450 nm. Both of these shift to smaller wavelengths with reduction in the size of the nanoparticle. These considerations show that for a nanoparticle in the range of 5-30 nm radius, the $l=1$ (dipolar mode) is the predominant accepting mode for energy transfer with a dye emitting in the visible range from 500-600 nm (absorption spectrum of gold nanoparticles lie in this range).

**3.1. Distance Dependence of the Rate of Energy Transfer:** The rate of energy transfer is calculated for both: the full Coulombic interaction and the dipole approximation approach using Eq. (2) where $H_I$ is given by Eq. (8) and Eq. (9) respectively. Fig. 5 illustrates that the rate of energy transfer ($k_{DA}$) is Förster type $(1/d^6)$ at large separations compared to the radius of the nanoparticle. However, at small separations ($d < 20a$), it breaks down and at distances approximately $d=a$ to $d=4a$, the rate varies as $(1/d^\sigma)$ where $\sigma$ lies between 3 and 4. In a recent experimental study, Strouse *et al.* found that the rate of energy transfer from a dye, a FAM moiety, to an Au-nanoparticle (diameter = 1.4 nm) can be fitted to a $(1/d^4)$ distance dependence.[27] These authors have suggested that this result may be understood in terms of surface energy transfer (SET).[20, 22] However these theoretical studies are mainly for the interaction of a dye with a *large* metallic surface - surface of a metallic "half-space". Though at intermediate distances we find $(1/d^4)$ but asymptotically distance dependence is still Förster type.

**3.2. Orientation Dependence of the Rate of Energy Transfer:** Because of the spherical symmetry of the nanoparticle, the orientation dependence is markedly different from that in two-dye system. In the latter case, depending on angle $\theta$ (defined as the angle between $\hat{R}$ and dye molecule, both being in the same plane), the normalized rate ($k_{DA}/k_{DA[max]}$) varies from 0 to 1.[19] If the dyes are oriented perpendicular to each other with the dipole of one of them oriented along $\hat{R}$ (this corresponds to $\theta = 0°$ for one dye and $\theta = 90°$ for the other), then there is no energy transfer. On the other hand, when the dyes are parallel to each other, $k_{DA}/k_{DA[max]}$ is either 1 (both $\theta = 0°$) or 0.25 (both $\theta = 90°$). The scenario is different in case of nanoparticle-dye system. At large separations ($d \gg a$), where the dipolar interaction Eq. (9) is accurate, the orientation dependence of the rate of energy transfer is governed solely by the second term $(\vec{\mu}_D.\hat{d})(\vec{\mu}_A.\hat{d})$. Since, the matrix element of $\vec{\mu}_A$ is parallel to that of $\vec{\mu}_D$, it follows that the orientation dependence of the rate is completely determined by the angle between the donor dipole and the vector $\hat{d}$. Further, it follows that, in contrast to the conventional FRET, there is no orientation that forbids energy transfer, and at large separation the ratio of the largest rate of transfer to the smallest rate of transfer approaches 4. Interestingly, the orientation dependence becomes weaker at smaller distances (see Fig. 6).

**3.3. Dependence of Energy Transfer Rate on the Size of Nanoparticle:** The energy transfer rate from a nanoparticle to a given dye is governed by Coulombic overlap integral Eq. (3), the position (surface plasmon frequency) and width (inverse surface plasmon lifetime) of the absorption spectrum of the nanoparticle relative to those of the dye. For a given dye, all the three are, in general, functions of the nanoparticle size. We briefly discuss the size dependence at large separation distances ($d \gg a$). For large nanoparticles ($a$ ~7 nm) since the plasmon frequencies are, to a very good approximation, independent of the size. Therefore, the energy transfer rate at large distances for large nanoparticles is determined entirely by the Coulombic overlap integral which we find to be proportional to the volume of the particle. For small nanoparticles both the plasmon frequency and lifetime depend on the size of the particle, hence the overlap of the absorption spectrum of the particle with the emission spectrum of the dye also contributes towards the size dependence of the rate. We have not studied the plasmon lifetime (inverse width of the absorption spectrum) in this work. Approximating the width of the absorption spectrum to be size independent (size dependence of the absorption spectrum has been studied using a time dependent density functional theory, for example, in Ref. 30), we have calculated the size dependence of the transfer rate at various distances as a function of nanoparticle size (see Fig. 7). These results agree with the asymptotics discussed above. Moreover, we find, interestingly, that at small separation distances, the energy transfer rate can even be non-monotonic with respect to the particle size (for small particle sizes). A more detailed study including the size dependence of plasmon lifetimes is necessary to uncover the complete picture.

In brief, the present discussion addresses the important issues of distance and orientation dependence of the rate of excitation energy transfer from a dye to a metal nanoparticle. The results presented here show that for most applications of FRET involving metal nanoparticle, the energy transfer shall involve surface plasmons and the asymptotic distance dependence remains Förster-type, although $1/d^6$ dependence breaks down at separations $< 20a$. The orientation factor varies from 1 to 4 as the dye molecule is rotated along the dye-nanoparticle axis from the perpendicular to the parallel orientation. The formalism adapted predicts an asymptotic $a^3$ size dependence of the rate of energy transfer. The present formalism can be easily extended to address the problem of energy transfer between two nanoparticles of different sizes or different metals as is discussed in the next section. The details of the study discussed in present section can be found in ref. 28.

## 4. Excitation energy transfer between two metal nanoparticles

The use of nanoparticles both as donor and acceptor in resonance energy transfer significantly increases the range over which the separations can be monitored. These RET systems are popularly referred to as 'plasmon rulers'.[10] The rate of energy transfer from one nanoparticle to the other depends not only on the separation between the two but also on the size and shape of the particles. In case of spherical particles the rate does not depend on the relative orientation of the nanoparticles. A recent experimental study demonstrated the plasmon coupling can be used to monitor separations of up to 70 nm between single pairs of gold and silver nanoparticles in vitro.[10]

Here we report the results of our study on the rate of energy transfer between two nanoparticles based on the formalism discussed above. The full Coulombic interaction and the dipolar interaction Hamiltonian is again given by Eq. (8) and Eq. (9) respectively. However, in present case the donor and acceptor integrals in full Coulombic interaction Hamiltonian are three dimensional.

Fig. 8 shows the schematic representation of the system under study. For resonance energy transfer to take place we need to consider two different size of nanoparticle with acceptor being larger in size than the donor. We consider the donor to be in first excited state corresponding to $l = 1$ mode while acceptor to be in the ground state i.e. no plasmon excitations. The distance dependence of the calculated rate is shown in Fig. 9. We find that for an acceptor size of 2 nm, the rate of energy transfer in case of two nanoparticle system is greater than that for a dye-nanoparticle system. As a result the large separations can be monitored with the former RET system. As discussed the rate of enegy transfer also depends on the size and shape of the nanoparticles. Though the qualitative dependence of the rate on distance will not change with the increase in the size of the nanoparticles but quantitative behaviour will definitely change. The further details of the study will be reported elsewhere.[31]

## Conclusion

The success of RET as a spectroscopic ruler depends critically on our knowledge of the distance and the orientation dependence of the rate of energy transfer. The present study involving nanoparticle reveals that while asymptotically we do have a Förster type $1/d^6$ distance dependence, at short separations comparable to the size ($a$) and even for somewhat larger separations, the rate varies as $1/d^\sigma$ with $\sigma$ varying from 3-4. Also for two conjugated dye molecules the deviation from $1/d^6$ has been observed. Note that $d$ in case of dye and nanoparticle system refer to the distance from the surface to the center of the dye molecule while for two nanoparticle system it is surface to surface distance. We find that unlike in conventional FRET the ratio of rate *( $k / k_{max}$ )* varies from 1 to 4 as the dye molecule is rotated along the dye-nanoparticle axis from the perpendicular to the parallel orientation. The formalism adapted predicts an asymptotic $a^3$ size dependence of the rate of energy transfer. We find that the range of separations that can be monitored substantially increases when "plasmon ruler" is employed. The rate also depends on the shape of the nanoparticle. The present study ignores the effects of vibrational relaxation in dye and also the effects of electron dynamics. These effects will result in broadening of lineshapes which has been introduced here as an approximation. In future, we hope to extend the theory to explicitly take account of these broadening factors. Work in this direction is under progress.

## Acknowledgments

It is a pleasure to thank Dr. Kim Wong, Prof. Peter Rossky and Prof. Harjinder Singh for collaboration and discussions. This work has been supported in part by grants from DST and CSIR. One of the authors (S.S.) acknowledges CSIR, India for research fellowship.

# FIGURE CAPTIONS

**FIG.1** Schematic representation of donor chromophore $PF_6$ and the acceptor TPP in an arrangement where the transition dipole moments are aligned (1) parallel to each other and orthogonal to DA intermolecular axis (cofacial parallel) and (2) parallel to each other and to the DA intermolecular axis (collinear parallel).z-axis shows the direction of the transition dipole moment vector.

**FIG.2** Distance dependence of the rate for the cofacial parallel orientation of donor and acceptor ((□) Förster rate and (•) resonance-Coulomb rates) for EET between the donor state and the acceptor state having mid-range oscillator strength. The traditional $R^{-6}$ distance dependence is shown by the solid line, the total Förster and resonance-Coulomb rates, summed over states are represented by dotted and dashed lines, respectively.

**FIG.3** Orientation dependence of normalized rate at **(a)** short (10Å) and **(b)** long (100Å) DA separation for an initial cofacial parallel alignment of DA transition moments ((□) Förster rate and (•) resonance-Coulomb rates).

**FIG.4** A schematic illustration of the geometric arrangement of the spherical nanoparticle and the dye molecule in two different orientations, parallel and perpendicular, with respect to the distance vector **R**. **d** is the distance measured from the surface of the nanoparticle. The figure also shows the coordinate system employed in our calculations.

**FIG.5** The distance (*d*) dependence of the rate of energy transfer ($k_{DA}$) calculated using full Coulomb interaction (solid line) and the dipole-dipole approximation (Förster theory) (dashed line) for the parallel orientation for a nanoparticle of size (*a*) 3 nm.

**FIG.6** Dependence of the rate of energy transfer on the orientation of the dye dipole moment $\vec{\mu}_D$ with respect to $\hat{d}$ [unit vector corresponding to $\hat{d}$ (see Fig. 4)]. The result

shown is for a gold nanoparticle of radius 1 nm, calculated using the full Coulombic interaction Eq. (8)

**FIG.7** Energy transfer rate as a function of the radius (*a*) of the nanoparticle and the distance (*d*) between the nanoparticle and the dye.

**FIG.8** A schematic illustration of the RET system involving two nanoparticles. The figure also shows the coordinate system employed in calculation.

**FIG.9** The distance (*d*) dependence of the rate of energy transfer between two nanoparticle. The radius of the donor nanoparticle is taken to be 1.5 nm while that for the acceptor to be 2 nm. The distance (*d*) is scaled with respect to the radius (*a*) of the acceptor. Note that the increase in size of the nanoparticles will not change the qualitative dependence of the rate on the distance.

# FIGURES

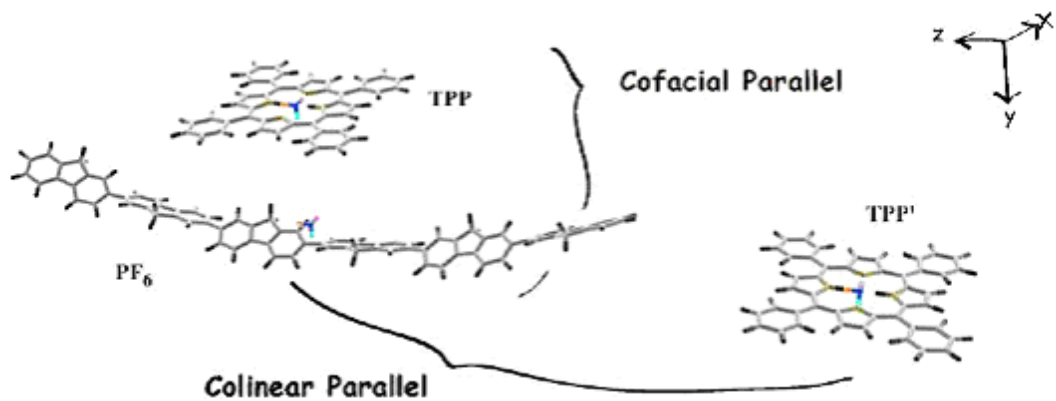

**FIG. 1**

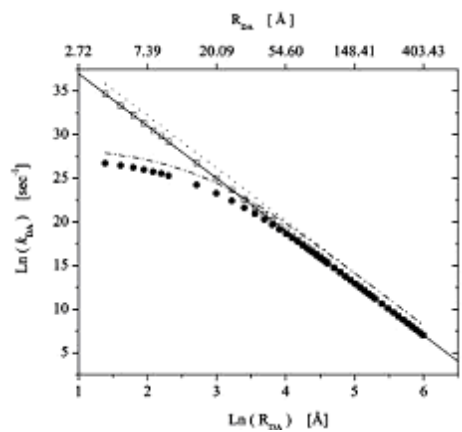

**FIG. 2**

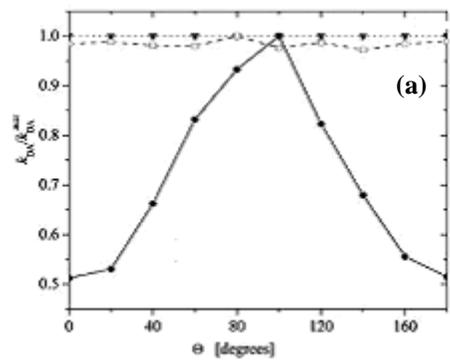 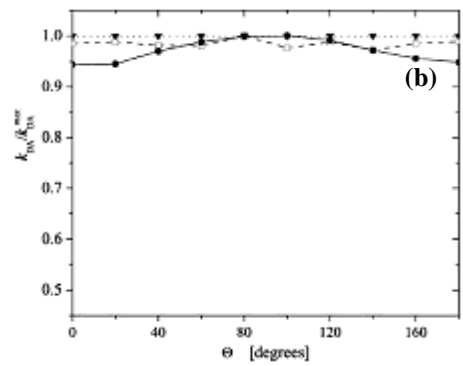

**FIG. 3**

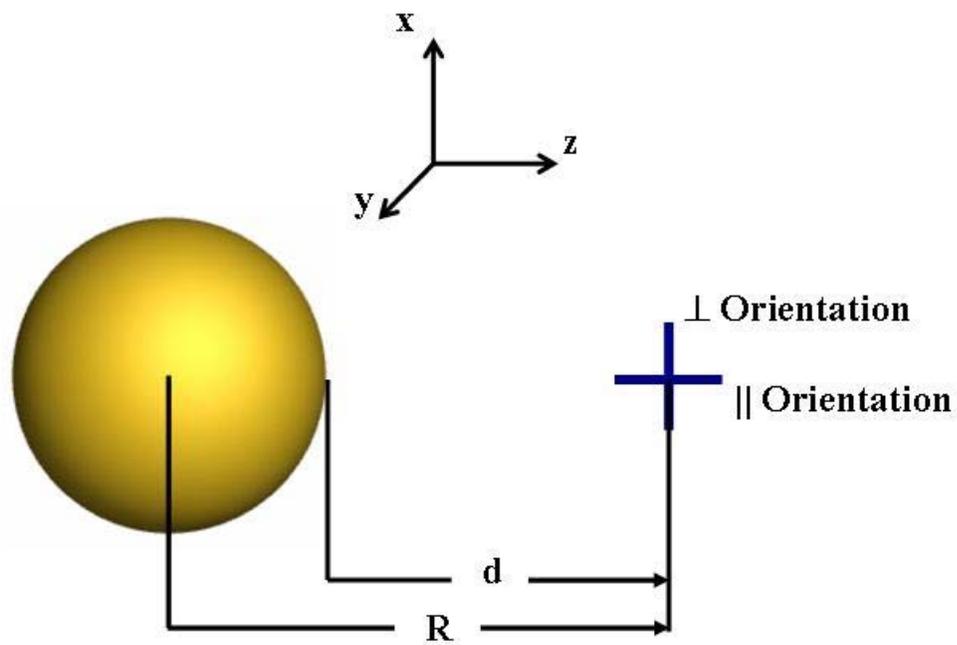

**FIG. 4**

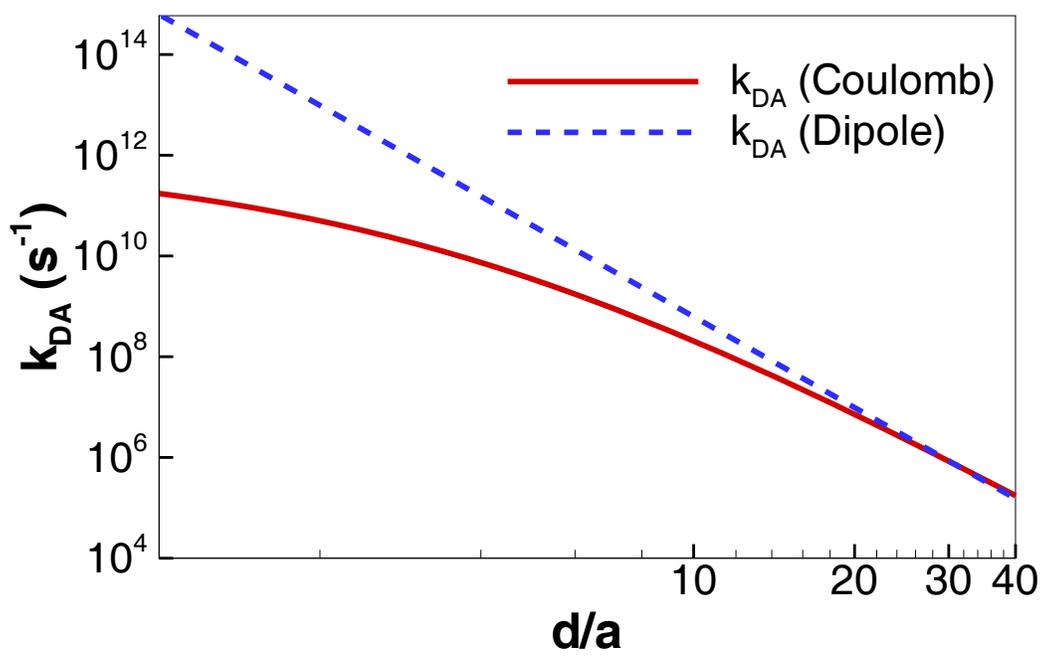

**FIG. 5**

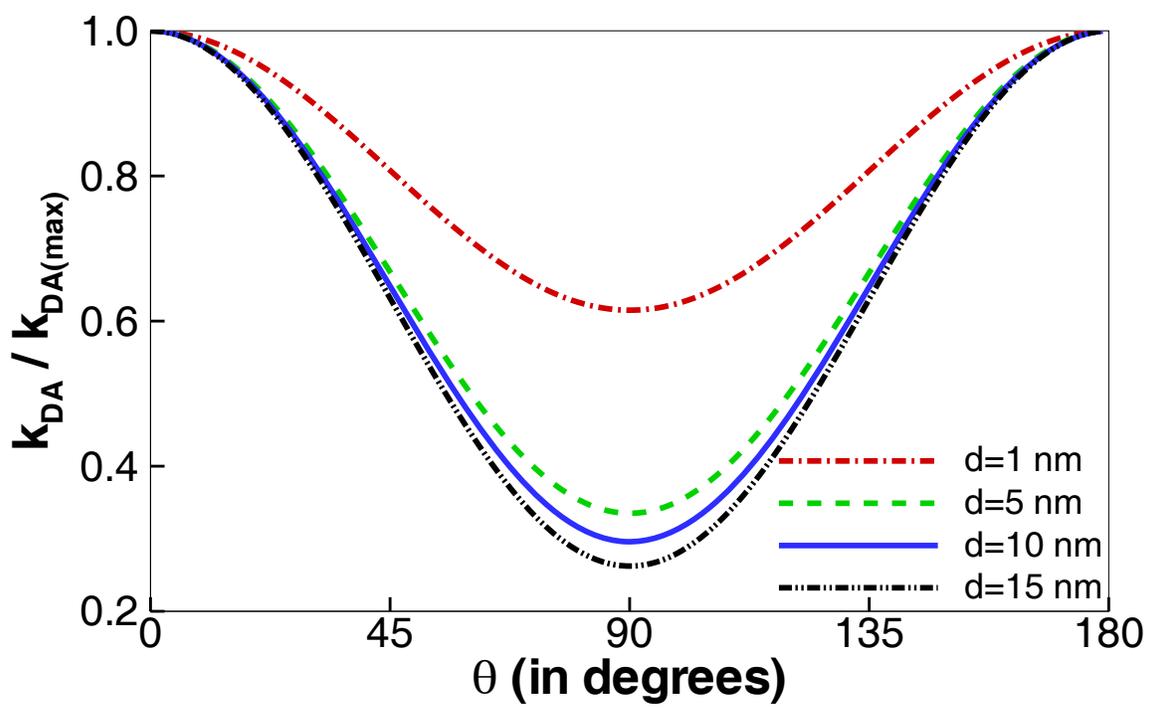

**FIG. 6**

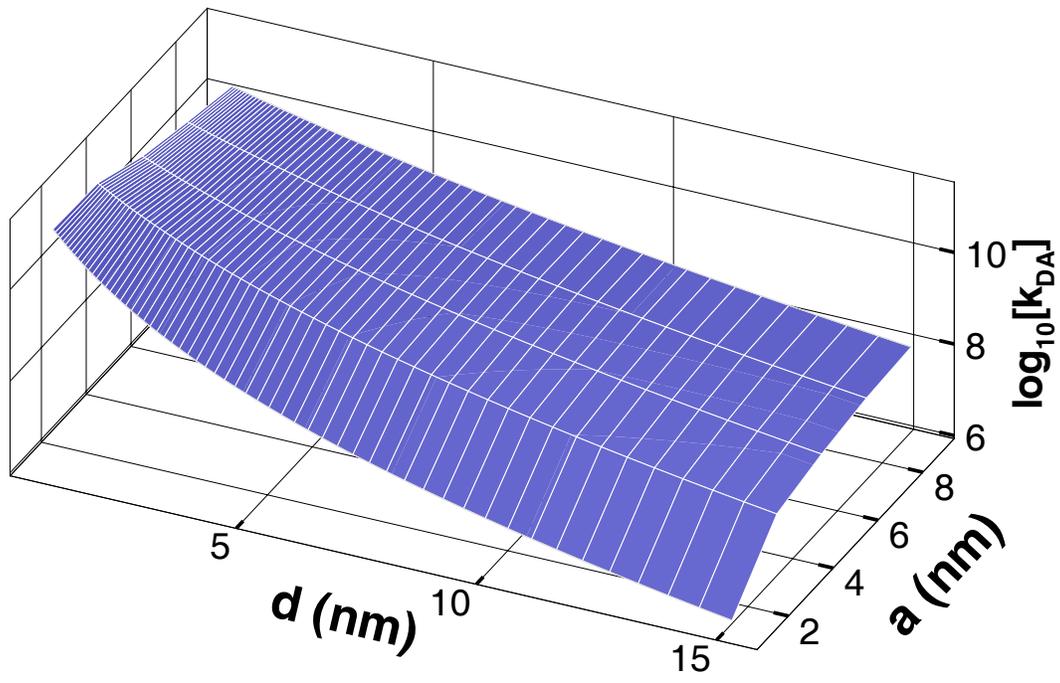

**FIG. 7**

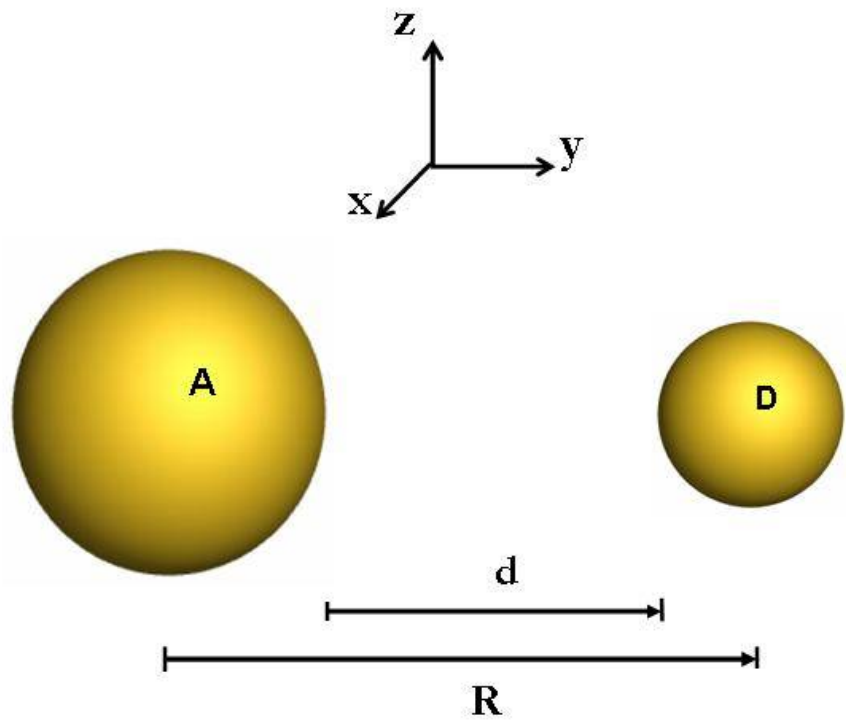

**FIG. 8**

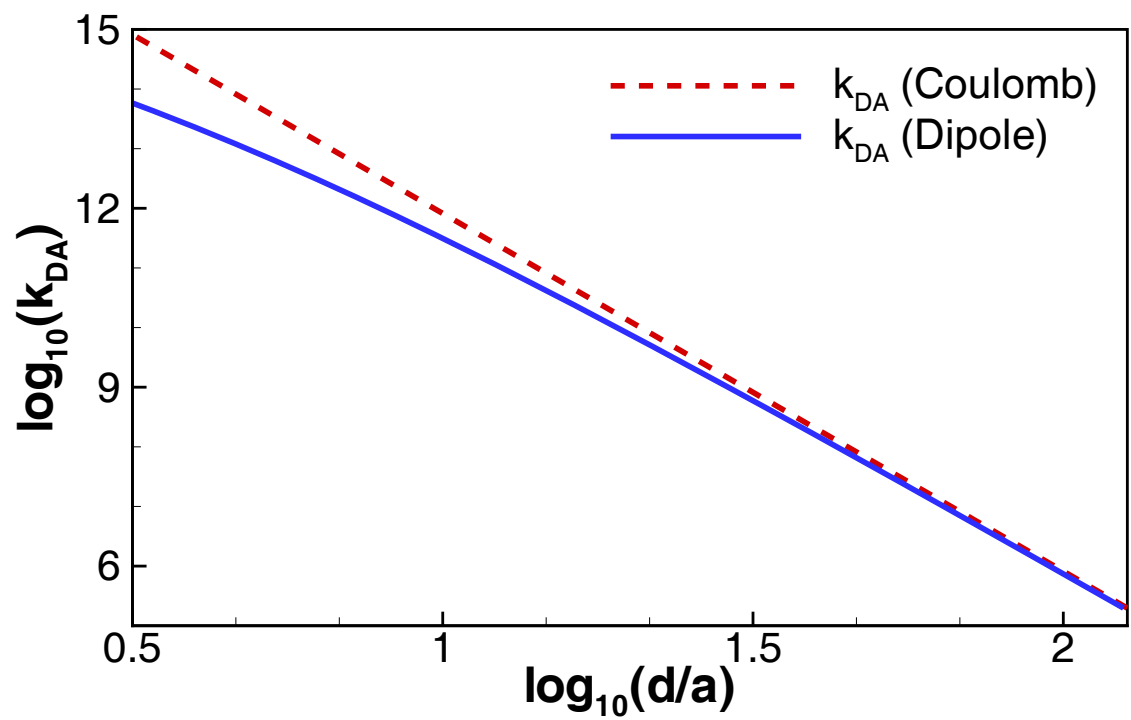

**FIG. 9**